\documentclass[aps,final,notitlepage,oneside,twocolumn,nobibnotes,nofootinbib,
superscriptaddress,noshowpacs,centertags]{revtex4-1}
\pdfoutput=1 % for breaklinks in bibliography with pdflatex in arxiv, the figures must be in PDF format

\usepackage[utf8]{inputenc}
\usepackage[english]{babel}
\usepackage{graphicx}
\usepackage[dvipsnames]{xcolor}
\usepackage[colorlinks,urlcolor=Blue,citecolor=Blue,linkcolor=Blue,pdfusetitle,breaklinks]{hyperref}
\hypersetup{breaklinks=true,colorlinks=true}
\usepackage{latexsym}
\usepackage{amssymb}
\usepackage{amsmath}
\usepackage{float}
\usepackage{color}
\usepackage{soul} 
\usepackage{natbib}
\usepackage[caption=false]{subfig}

\usepackage{booktabs}
\usepackage{lipsum}

\newcommand{\pbh}{{\rm PBH}}
\newcommand{\eq}{{\rm eq}}

\newcommand{\crit}{{\rm crit}}

\begin{document}

\title{Merger History of Clustered Primordial Black Holes}

\author{Viktor Stasenko}\email{vdstasenko@mephi.ru}
\affiliation{National Research Nuclear University MEPhI, 115409 Moscow, Russia}
\affiliation{MIREA --- Russian Technological University, 119454 Moscow, Russia}

\begin{abstract}
    Primordial black hole (PBH) binaries experience strong gravitational perturbations in the case of their initial clustering, which significantly affects the dynamics of their mergers. In this work, we develop a new formalism to account for these perturbations and track the evolution of the binary orbital parameters distribution. Based on this approach, we calculate the merger rate of PBH binaries and demonstrate that its temporal evolution differs greatly from that of isolated binary systems. Moreover, PBH clustering produces distinctive features in the stochastic gravitational-wave background: the canonical $2/3$ spectral slope transforms to $\Omega_{\rm gw} \propto \nu^{-65/28}$ in a certain frequency band. These predictions can be probed in future gravitational wave observations, opening up new opportunities to test the clustering of PBHs and their contribution to dark matter.
\end{abstract}

\maketitle

\section{Introduction}

Progress in gravitational-wave astronomy has led to a surge of interest in primordial black holes (PBHs) --- hypothetical black holes that may have formed in the very early Universe.~\cite{1967SvA....10..602Z, 1971MNRAS.152...75H}. In recent years, the LIGO-Virgo-KAGRA collaboration has detected dozens of compact object merger events~\cite{KAGRA:2021duu,KAGRA:2021vkt}. Most of them are interpreted as mergers of black holes of astrophysical origin, but the primordial nature of a part of the detected events is also popular~\cite{Bird:2016dcv,Sasaki:2016jop,Clesse:2016vqa,Blinnikov:2016bxu,Ali-Haimoud:2017rtz,Raidal:2018bbj,Liu:2018ess,DeLuca:2020qqa,Hutsi:2020sol,Fakhry:2020plg,Fakhry:2021tzk,Fakhry:2023fbt}. This interpretation is particularly valuable because it allows us to use gravitational-wave (GW) observations as a tool to impose constraints on the density of PBHs in the Universe. Moreover, many other independent observations have imposed additional constraints, but primordial black holes remain a viable candidate for dark matter~\cite{Carr:2020gox} --- one of the central problems of present-day fundamental science.

It is traditionally considered that PBH binaries are formed at the radiation-dominated stage of the Universe evolution due to fluctuations in their spatial distribution~\cite{Nakamura:1997sm,Sasaki:2016jop}. Despite a well theoretical elaboration of this scenario, quantitative estimates of the merger rate face uncertainties. The main difficulties are related to various factors that affect the binary systems: in particular, the interaction of the binaries with the dark matter halo that forms around them~\cite{Kavanagh:2018ggo,Pilipenko:2022emp,Jangra:2023mqp,Stasenko:2024dui}; perturbations of PBH binaries in early structures arising from their natural Poisson clustering~\cite{Raidal:2018bbj,Vaskonen:2019jpv,Jedamzik:2020ypm,Tkachev:2020uin,Stasenko:2023zmf,Stasenko:2024pzd,Delos:2024poq}; and other possible effects. 

In this paper, we consider an alternative scenario assuming an initially clustered birth of PBHs. In this case, the statistics of PBH binaries is formally constructed similarly to the model of their random distribution in space~\cite{Nakamura:1997sm,Sasaki:2016jop}. However, in the case of strong clustering, the key difference is that the PBH binaries appear to be embedded in a dense environment. In such conditions they are subjected to continuous gravitational perturbations from the surrounding single PBHs. In this paper we develop a new formalism describing how these perturbations affect mergers of PBH binaries. This aspect has not been considered in previous papers~\cite{Bringmann:2018mxj, Young:2019gfc,Atal:2020igj,Crescimbeni:2025ywm} examining the impact of initial clustering on PBH mergers. Our calculations show that this process leads to a significantly different temporal evolution of the merger rate compared to the prediction for PBHs without initial clustering. In particular, intense perturbations can both destroy the existing binary systems and significantly modify the orbital parameters of the surviving binaries, increasing their lifetime. These effects have a significant impact on the spectral characteristics of the stochastic GW background, which integrates the contribution of black hole mergers over cosmological history to the GW energy density in the Universe. The results have important observational implications for next generation of GW experiments such as LISA~\cite{amaro2017laser}, TianQin~\cite{TianQin:2015yph}, Einstein Telescope~\cite{ET:2019dnz,Abac:2025saz} and other proposed projects. Their expected sensitivity will allow both the detection of individual black hole merger events at high redshifts and the measurement of the GW background over a wide frequency range. This will provide a test of the PBH idea that was discussed in a series of papers, for example in Refs.~\cite{Ng:2022agi,Ng:2022vbz,Branchesi:2023mws, Franciolini:2023opt,LISACosmologyWorkingGroup:2023njw,Abac:2025saz}.

The paper is organized as follows. Section~\ref{sec2} develops the simple theoretical formalism for PBH cluster formation. In Section~\ref{sec3}, we analyze the formation of PBH binaries during the radiation-dominated era of the Universe. Section~\ref{sec4} provides both qualitative analysis of the perturbation processes affecting binary systems and analytical estimates for the parameters of binaries undergoing perturbations. Section~\ref{sec5} presents our calculations of the PBH merger rate incorporating these perturbation effects. The implications of our results for the stochastic GW background are discussed in Section~\ref{sec6}. Finally, Section~\ref{sec7} summarizes our main findings.

\section{Primordial black hole clustering}\label{sec2}

A number of theoretical models predict that the formation of PBHs is accompanied by their strong spatial clustering, which has been shown, for example, in the works~\cite{Rubin:2001yw, Khlopov:2004sc, Young:2015kda, Tada:2015noa, Belotsky:2018wph, Desjacques:2018wuu, Suyama:2019cst, Ding:2019tjk}. In this case, there must exist regions of some spatial scale with nonzero PBH density and the abundance of such clusters in the Universe is determined by the relative density of the PBHs $\Omega_\pbh$. To parametrize the inhomogeneous PBH distribution and account for their locally density enhancements, we introduce the density contrast $\delta_\pbh (\bf x) = \rho_{\rm PBH}(\bf x) / \rho_{\rm DM}$, where $\rho_{\rm DM}$ is the background dark matter density. The quantity $\delta_\pbh$ is model-dependent, but in this work we consider it as free parameter, the specific choice of which determines the moment of gravitational detachment of the cluster from the Hubble flow. Although the quantity $\delta_\pbh$ should generally vary across spatial scales, we ignore its scale dependence (power spectrum) since our goal is to show the influence of clustering on PBHs mergers. Therefore we assume clusters have a monochromatic mass function, and in our simplified model $\delta_\pbh$ is nonzero only within some small spatial region of comoving size $R\leq R_{\rm cl}$ from the cluster center: 
\begin{equation} \label{eq1}
    \delta_\pbh(R) = 
    \begin{cases}
        \delta_\pbh & \text{if } R \leq R_{\rm cl}, \\
        0 & \text{otherwise}.
    \end{cases}
\end{equation}
Moreover $\delta_\pbh$ affects the velocity dispersion in the cluster and we will discuss this at the end of this Section. We parametrize the fraction of PBHs in the dark matter composition in the traditional way $f = \Omega_{\rm PBH}/\Omega_{\rm DM}$. Moreover, for clustering $\delta_\pbh > f$ must be fulfilled, the case $\delta_\pbh = f$ corresponds to the homogeneous distribution of PBHs in the Universe.

The formation of gravitationally bound structures occurs when the relative density contrast reaches a value on the order of unity $\delta \rho/\rho \sim1$. In the radiation-dominated epoch, the radiation density falls as $\rho_{r} =\rho_{\rm eq} (s/s_{\rm eq})^{-4}$, and the density of PBH (like dark matter) $\rho_{\rm PBH} = \rho_{\rm eq} \delta_\pbh (s/s_\eq)^{-3}$, where $s_{\eq}$ is the scale factor at the moment of matter-radiation equality $t_{\rm eq}$. At the present moment, the scale factor is normalized to unit $s_0 = 1$. Regions with locally enhanced PBH density are isocurvature fluctuations that remain ``frozen'' during the radiation-dominated era, so the moment of formation can be determined from the condition $\rho_{r} \sim \rho_{\rm{PBH}}$. Hence we obtain an estimate for the moment of cluster formation $\delta_\pbh \sim s_{\rm eq}/s_f$. In this way, a gravitationally bound structure (cluster) is formed, separated from the Hubble expansion, consisting of PBHs and dark matter particles with density $\rho_{\rm cl} = B \rho_{\eq} \delta_\pbh^3(1 + \delta_\pbh)$, where the constant $B$ is determined from the numerical solution of this problem and will be given below. In the case $\delta_\pbh < 1$, cluster formation occurs at the dust stage of the Universe, when structures form due to gravitational instability. During this period, small density fluctuations grow in the linear regime as $\delta \rho /\rho \propto s$, which leads to a similar estimate for the moment of cluster formation $\delta_\pbh \sim s_{\rm eq}/s_f$. The analysis of~\cite{Kolb:1994fi} demonstrates that in both cases the density is well described by a simple expression
\begin{equation}\label{eq:rho_cl}
    \rho_{\rm cl} \approx 140 \, \rho_{\rm eq}  \delta_\pbh^3 \big(1 + \delta_\pbh \big).
\end{equation}
The density of the PBHs in the cluster will be $\rho_{\rm PBH} \approx 140 \, \delta_\pbh^4 \rho_{\rm eq}$. Here $\rho_{\eq} = \Omega_{\rm DM} \rho_{\rm crit} (1 + z_{\eq})^3$, $z_{\rm eq} = 3402$ is the redshift matter-radiation equality and $\rho_{\rm crit} = 3 H_0^2 /8 \pi G$ is the critical density of the Universe. The expression for the Hubble parameter is 
\begin{equation}
    H^2(z) = H_0^2 \Big [\Omega_{\Lambda} + \Omega_M (1+z)^3 + \Omega_r (1+z)^4 \Big].
\end{equation}
In our calculations we use the cosmological parameters $H_0 = 67.4$~km~s$^{-1}$~Mpc$^{-1}$, $\Omega_{\Lambda} = 0.69$, $\Omega_M = 0.31$, $\Omega_r = 5.4 \times 10^{-5}$~\cite{Planck:2018vyg}. 

Let us now discuss the spatial scale of clustering. Our consideration assumes that PBH clusters are independently distributed in space. Then the clusters obey Poisson statistics as point-like objects, therefore the clusters are subject to the Lyman$-\alpha$ forest constraints, which implies  $f Nm \lesssim 100 \,M_{\odot}$~\cite{Murgia:2019duy}, here $N$ and $m$ are the number and mass of the PBHs in the cluster, respectively. For
definiteness we will assume that $f \simeq 0.01$, which is also consistent with recent microlensing constraints~\cite{Mroz:2024mse, Mroz:2025xbl}. Then the total mass of PBHs in the cluster should be $N m \lesssim 10^4 \, M_{\odot}$. The total mass of the cluster also includes dark matter $M_{\rm cl} = Nm(1 + \delta_\pbh)/\delta_\pbh$, and using Eq.~\eqref{eq:rho_cl} we obtain the radius of the cluster
\begin{equation}
    r_{\rm cl} \sim \left ( \frac{Nm}{140 \, \rho_{\eq} \delta_\pbh^4} \right)^{1/3} \approx \frac{0.3 \, \rm{pc}}{\delta_\pbh^{4/3}} \left ( \frac{Nm}{10^4 \, M_{\odot}} \right)^{1/3}. 
\end{equation}
The comoving scale of the cluster will be
\begin{equation}
    R_{\rm cl} \sim \left ( \frac{Nm}{\rho_{\rm DM,0}} \right)^{1/3} \approx 6.7 \left ( \frac{Nm}{10^4 \, M_{\odot}} \right)^{1/3} \, \rm{kpc}, 
\end{equation}
where $\rho_{\rm DM,0} \approx 33\, M_{\odot}$~kpc$^{-3}$ is the present-day dark matter density. The scales under consideration are much smaller than those observed by the Planck satellite, at which constraints are imposed on the isocurvature modes. Thus, PBH clusters make up only a small part of the dark matter bulk, and their physical size are much smaller than the distance between them $r_{\rm cl} \ll R_{\rm cl}/f^{1/3}$. 

In this work we are interested in the gravitational dynamics of the PBHs with each other, so the important parameter for us is the velocity dispersion of PBHs in the cluster which is estimated as 
\begin{equation}
    \sigma \sim \sqrt{\frac{GM_{\rm cl}}{r_{\rm cl}}} \approx 12 \, \big  (1 + \delta_\pbh\big)^{1/2} \left (\frac{\delta_{\pbh}^{1/2} Nm}{10^4 \,M_{\odot}}\right)^{1/3} \rm{km\,s^{-1}}. 
\end{equation}
Our results will be presented as a function of velocity dispersion $\sigma$, allowing straightforward reconstruction of their dependence on spatial scale $R_{\rm cl}$. All numerical calculations are carried out for PBHs with mass $m = 10\, M_{\odot}$, which is within the mass range of the registered LIGO-Virgo-KAGRA events. However we also provide analytical expressions of our results that allow us to generalize the findings to the case of arbitrary masses of the PBHs. Also, throughout the paper we adopt the approximation $\Omega_{\rm DM} = \Omega_{\rm M}$, since taking into account the baryonic component will only make minor corrections to our results. 

\section{Primordial black hole binaries} \label{sec3}

During the radiation-dominated stage, the PBHs form gravitationally bound binary systems. This occurs because at some moment of time $t_{\rm dec} < t_{\rm eq}$ the local density of two PBHs will exceed the radiation density $\rho_{r} (t_c)$. At the same time, these black holes decouple from the cosmological expansion and form the binary. In this sense, the situation is completely analogous to the formation of clusters discussed in the previous section. In this Section, we will assume that PBHs are uniformly distributed within the region with their increased density (in the cluster). Therefore, the results of Ref.~\cite{Sasaki:2016jop} are applicable to our consideration. First of all, we introduce the mean physical distance between the PBHs at the moment $t_{\rm eq}$
\begin{equation}
    \overline{x} = \left ( \frac{m}{\rho_{\rm{PBH}}}  \right)^{1/3} = \frac{1}{s_{\eq} \delta_{\pbh}^{1/3}} \left ( \frac{m}{\Omega_{\rm{DM}} \rho_{\crit}}  \right)^{1/3}.
\end{equation}
Since the radiation density falls as $\rho_r \propto s^{-4}$, the moment of gravitational decoupling of the two PBHs is determined by the following expression
\begin{equation} \label{dec}
	\frac{m}{R^3} = \rho_r(t_{\rm dec}) = \rho_{r,\rm eq} \left ( \frac{s_{\eq}}{s_{\rm dec}} \right)^4,
\end{equation}
here $R = x s_{\rm dec}/s_{\eq}$ is the physical distance between the PBHs at the moment $t_{\rm dec}$, and $x$ is the distance at the moment $t_{\rm eq}$. Thus, the moment of formation of the binary system is determined as 
\begin{equation}
    \frac{s_{\rm eq}}{s_{\rm dec}} = \delta_{\pbh} \left ( \frac{\overline{x}}{x}\right)^{3}.
\end{equation}
In the described formalism, the decoupling moment has two natural bounds: $s_{\rm dec} < s_\eq$ in the case $\delta_{\pbh} < 1$ (similar unclustered scenario) and $s_{\rm dec} < s_\eq/\delta_{\pbh}$ when $\delta_{\pbh} > 1$. The second condition arises because the gravitational detachment of the PBH pair from the Hubble flow must occur before the formation of the cluster. As was shown in Sec.~\ref{sec2}, the moment of PBHs cluster formation is given by $s_f \sim s_\eq/\delta_\pbh$, and the formation of a PBH binary must precede the formation of the cluster $s_{\rm dec} < s_f$, as the reverse is not physically sensible. Taking this into account, as well as the fact that $x < \overline{x}$, we conclude that binaries are formed only by those PBHs whose distance between them satisfies the relation $x < \overline{x} \delta_\pbh^{\gamma}$, where $\gamma$ is used to parametrize two possible clustering modes: $\gamma = 1/3$ for $\delta_\pbh < 1$ and $\gamma = 0$ for the case $\delta_\pbh > 1$. 

After detachment from the expansion of the Universe, the PBHs avoid a head-on collision and move in an elliptical orbit around each other because tidal torques from neighboring PBHs induce angular momentum. Here, for simplicity, we consider only the main contribution from the nearest PBH, the distance to which we denote by $y$, and the condition $x < y < \overline{x}$ must be satisfied. We continue to follow the formalism of work~\cite{Sasaki:2016jop} and take the probability distribution over $x$ and $y$ to be flat in three-dimensional space 
\begin{equation} \label{eq:dp_dxdy}
    dP = \frac{9}{\overline{x}^6} x^2 y^2 \, dx dy. 
\end{equation}
The major and minor axes of the PBH binary are determined by the equations
\begin{equation} \label{eq:axby}
    a = \frac{\alpha x}{\delta_\pbh} \left ( \frac{x}{\overline{x}} \right)^3, \quad b = \beta a \left ( \frac{x}{y} \right)^3,
\end{equation}
further, the coefficients $\alpha$ and $\beta$ are taken to be equal to unity. The eccentricity of a binary is defined as $e =\sqrt{1 - b^2/a^2}$, but it is more convenient to use the dimensionless angular momentum $j = \sqrt{1 - e^2} = (x/y)^3$ instead. To characterize the PBH binaries, it is suitable to transform to the variables $a$ and $j$ in the distribution \eqref{eq:dp_dxdy}  
\begin{equation} \label{eq:dp_dadj}
    dP = \frac{3}{4} \left ( \frac{\delta_\pbh}{\overline{x}} \right)^{3/2} \frac{\sqrt{a}}{j^2} \, da dj,
\end{equation}
where the variables vary in the range $0<a<\overline{x} \delta_\pbh^{4\gamma-1}$, $(a \delta_\pbh/\overline{x})^{3/4} < j < 1$. The lower bound on $j$ corresponds to the fact that the distance to the third PBH is maximized $y = \overline{x}$. From the distribution of~\eqref{eq:dp_dadj} we can see that PBH binaries mainly have small angular momenta at formation, i.e., their orbits are strongly elongated. Due to the emission of GW a binary black holes of the same mass $m$ merge in time~\cite{PhysRev.136.B1224}
\begin{equation} \label{eq:tgw}
    t_{\rm gw}=\frac{3 c^5 a^4 j^7}{170 G^3m^3}.
\end{equation}

Let us now determine the characteristic initial values of the parameters of the PBH binaries merging by a time $t_{\rm mer}$. As noted above, the distribution~\eqref{eq:dp_dadj} is maximal at small $j$, so the characteristic value of the angular momentum can be estimated as $j_{\rm ch} \sim j_{\rm min} = (a \delta_\pbh/\overline{x})^{3/4}$. Then from the binary merger time~\eqref{eq:tgw} we obtain the characteristic value for the major semiaxis 
\begin{align} \label{eq:ach}
    a_{\rm ch} &= \left ( \frac{170 G^3m^3t_{\rm mer}}{3c^5} \right)^{4/37} \left ( \frac{\overline{x}}{\delta_\pbh} \right)^{21/37} \nonumber \\
    &\approx 42 \, \delta_\pbh^{-28/37} \left ( \frac{m}{M_{\odot}} \right)^{19/37} \left ( \frac{t_{\rm mer}}{t_0} \right)^{4/37} \, \rm{au},
\end{align}
where $t_0 =13.8$~Gyr is the age of the Universe. Since $a_{\rm ch}$ depends very weakly on $t_{\rm mer}$, almost all merging binaries at not too far cosmological distances are characterized by approximately the same value of the large semi-axis. Now let us compare the obtained value with the maximum possible major semiaxis. In the case of $\delta_\pbh < 1$, the limiting value of $a_{\rm max}$ corresponds to the average distance between the PBHs at $\delta_\pbh =1$, which is $a_{\rm max} \sim 10^4$~au. In the case $\delta_\pbh > 1$, we get $a_{\rm max} \sim 10^4 \delta_\pbh^{-4/3}$~au. Thus, all PBH binaries merging by the current epoch formed gravitationally bound pairs long before the transition to the stage of matter dominance (or before the formation of the cluster $t_f$ at $\delta_\pbh > 1$). Substituting the found expression for the  semi-major axis~\eqref{eq:ach} into $j_{\rm min} = (a \delta_\pbh/\overline{x})^{3/4}$ one can determine the characteristic value of the angular momentum of the binaries
\begin{equation} \label{eq:jch}
    j_{\rm ch} \approx 0.01 \, \delta_{\rm PBH}^{16/37} \left (\frac{m}{M_{\odot}} \right)^{5/37} \left ( \frac{t_{\rm mer}}{t_0} \right)^{3/37}.
\end{equation}
Thus, PBH binaries are characterized by small angular momenta $j \ll 1$, which will be used later in the analysis of the interaction of such binaries with single PBHs. 

\section{Perturbations of binaries} \label{sec4} 

Since PBH binaries are embedded in clusters, they are subjected to gravitational perturbations by single black holes. Because of this, the orbital parameters of the binaries change. The crucial parameter determining the outcome of such interactions is the hardness of the binary. A system is considered hard if its binding energy $E_b$ exceeds by absolute value the characteristic kinetic energy of black holes in the cluster $\sim m\sigma^2$. In terms of orbital parameters this condition corresponds to the constraint on the large semi-axis 
\begin{equation}\label{eq:ah}
    a < a_h = \frac{Gm}{2\, \sigma^2} \approx 4.4 \, \left ( \frac{\sigma}{10\,\rm km \, s^{-1}} \right)^{-2} \left ( \frac{m}{M_{\odot}}\right) \, \text{au},
\end{equation}
Binary systems with large semi-major axis $a > a_h$ are classified as soft, otherwise --- hard binaries. According to the classical Heggie-Hills law, such hard binaries tend to become harder when interacting with single objects ($|E_{b}|$ increases), while soft ones become softer~\cite{1975MNRAS.173..729H,1975AJ.....80..809H}.

The comparison of Eqs.~\eqref{eq:ach} and~\eqref{eq:ah} shows that when $\delta_\pbh \sim 1$ and the velocity dispersion $\sigma \sim 10$~km~s$^{-1}$ the characteristic binaries~\eqref{eq:ach} are soft, so they will become softer (or even destroy) as a result of perturbations. Moreover, since their angular momentum $j \ll 1$, it will mainly increase under the influence of perturbations. This statement is also true for hard binaries with large eccentricities. The physical reason for this behavior is that the angular momentum distribution from the initial state~\eqref{eq:dp_dadj} tends to the thermal one, for which $dP/dj \propto j$. The significant growth of the angular momentum was also confirmed by numerical modeling of such interactions in the work~\cite{Jedamzik:2020ypm}. The increase of the angular momentum, in turn, leads to a significant increase of the lifetime of the binary. In particular, the increase of the angular momentum from the characteristic value $j_{\rm ch}$ to $j\sim 1$ causes the merger time to increase by a factor of $\sim 10^{14}$. Therefore, binaries that form on highly eccentric orbits, under the influence of perturbations, increase the time to merger by many orders of magnitude. 

A different dynamics is expected for binary systems on nearly circular orbits. In their case, the perturbations will keep the angular momentum at $j \sim 1$. This makes such binaries ``stable'', i.e. their merger time is practically unchanged by perturbations. Although hard binaries due to perturbations should still become harder, with a corresponding decrease of the merger time. However, it is difficult to quantitatively account for this effect because the hard binaries are eventually ejected from the cluster. Although predicting the orbital parameter distribution of the ejected binaries is of interest, the bulk of the binaries in the cluster still have small angular momenta. Therefore, in our calculations we restrict ourselves to the approximation in which the near circular binaries with $j > j_{\rm cut}$ are considered stable: their orbital evolution and merger time are assumed to be unchanged. As a threshold value we take $j_{\rm cut} = 1/2$; as will be seen below, the variation of this parameter does not significantly affect the main results of our work.  

\begin{figure*}[t!]
	\centering
	%
	%\subfloat[name fig]
        \subfloat{
	\hspace*{-1.00cm}
	\includegraphics[width = 0.48\textwidth]{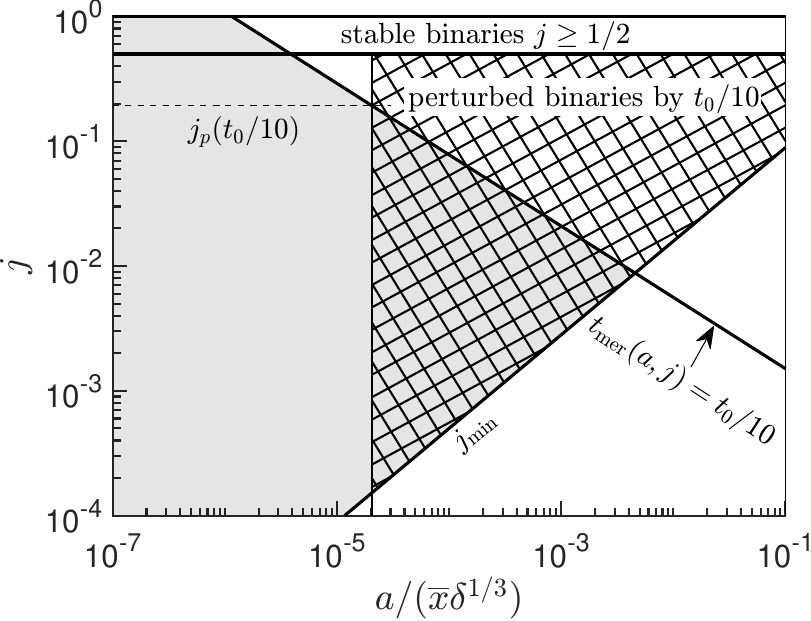}
	%\label{fig:area_jt}
        }
	\subfloat{
	\hspace*{-0.00cm}
	\includegraphics[width = 0.48\textwidth]{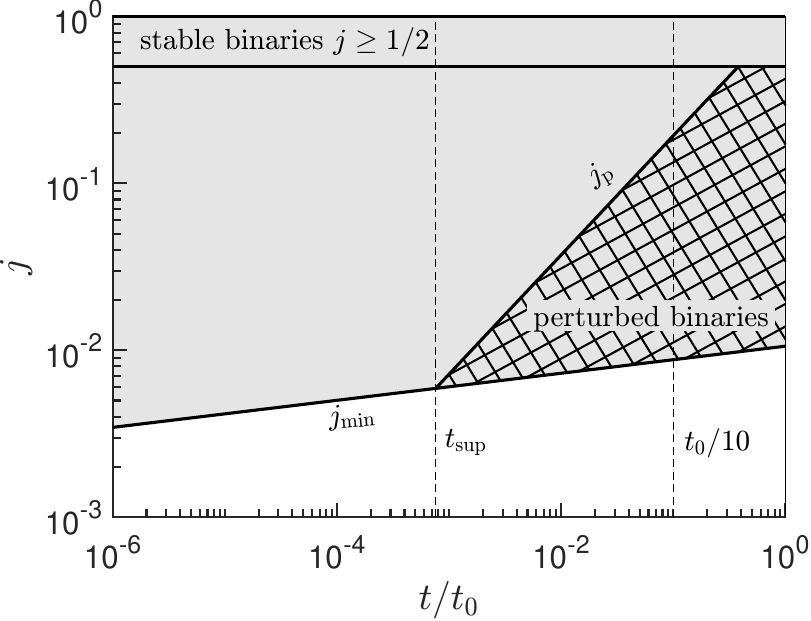}
        }
    \caption{The example of the orbital parameter space of PBH binaries with $\delta_\pbh =0.5$, $\sigma = 10$~km~s$^{-1}$ and $m = 10\, M_{\odot}$. \textit{Left}: Binaries merging by time $\leq t_0/10$ are shaded in gray. The hatched area (which emphasizes that binaries with these parameters ``drop out'' of the merger process) schematically shows binaries that have been perturbed at the same time, for which the semimajor axis exceeds the value given by $\tau(a) = t_0/10$. For reasons given in the text, perturbations of binaries with a nearly circular orbit ($j \geq 1/2$) are assumed not to affect their merger time. The dashed horizontal line indicates the critical angular momentum of binaries $j_p$~\eqref{eq:jp}, separating tight systems that avoid perturbation to $t_0/10$ and wide perturbed binaries.
    \textit{Right}: Similar to the left parameter space, but on the $(t,j)$ plane, where the transformation from $a$ to $t$ is done using Eq.~\eqref{eq:tgw}. At fixed time $t$, the merging binaries have angular momenta from $j_{\rm min}$ to $1$. However, systems with small angular momenta (corresponding to wide binaries) are perturbed, leading to an effective constraint from below on the value of $j_p$. 
	}
	\label{fig:area_b}
\end{figure*}

Let us now pass to the description of the perturbation of binary systems. The cross section of the approach of a binary, considered as a point object, and a single black hole to the distance $r_p$ (pericenter of the orbit) is defined by the expression 
\begin{equation} \label{eq:cs_h}
    \Sigma_{p} = \pi r_p^2 \left (1 + \frac{6 G m}{r_p v^2} \right),
\end{equation}
here $v$ is the relative velocity between the binary and single PBH at infinity. In our simplified model it is assumed that, as a result of a single act of hard scattering, the angular momentum of the binary increases to $j\sim 1$ and hence its merger time increases significantly. To define hard scattering we use a simple criterion, which is that this distance of minimum approaching is equal to the major semiaxis $r_p = a$. The probability of such interactions is determined by the mean free time $\tau^{-1} = \rho_{\pbh} \langle \Sigma_p v \rangle /m$. Assuming the Maxwell distribution of PBHs velocities in the cluster with dispersion $\sigma$, and after averaging over the relative velocities, we obtain the characteristic perturbation time of orbital parameters of PBH binaries
\begin{equation} \label{eq:tau}
    \tau^{-1} = \frac{2 \sqrt{2 \pi}}{m} 140 \, \rho_{\eq} \delta_\pbh^4 \sigma a^2 \left (1 + \frac{3\, Gm}{a \sigma^2} \right),
\end{equation}
where the density of black holes in the cluster is $\rho_{\rm PBH} \approx 140 \, \delta_\pbh^4 \rho_{\rm eq}$ (see Section~\ref{sec2}). First of all, let us estimate the perturbation time of the characteristic binaries~\eqref{eq:ach} merging in the current epoch. As an example, we consider the case of the PBH with mass $m = 10\,M_{\odot}$ with clustering parameter $\delta_\pbh = 1$ and velocity dispersion $\sigma = 10$~km~s$^{-1}$. The perturbation time is estimated to be $\tau \approx 0.6$~Myr, and since the semimajor axis of typical binaries~\eqref{eq:ach} depends weakly on the merger time, it can be concluded that typical binaries are subject to perturbation for almost the entire cosmological history after their formation. Let us now consider the case of binaries in a circular orbit and with coalescence time $t_0$. At the time of formation, the semimajor axis of such a binary should be $a \approx 0.8$~au, and their perturbation time turns out to be $\tau \approx 2$~Gyr. Note that for a cluster with velocity dispersion $\sigma > 75$~km~s$^{-1}$ these binaries turn out to be even soft. These estimates show that in the considered example all binaries that potentially produce observable GW signals are perturbed in clusters. In the case of earlier time scales $<t_0$ or with decreasing $\delta_\pbh$ and PBH masses, the perturbations affect only a certain fraction of all binaries, namely only sufficiently wide binaries are subject to perturbations. 

Let us now consider binaries with some merger time $t_{\rm mer}$. Using Eq.~\eqref{eq:tgw}, we express the major semiaxis through $t_{\rm mer}$ and angular momentum $j$ to obtain the perturbation time $\tau(t_{\rm mer}, j)$. The binaries with $\tau > t_{\rm mer}$ merge before a significant perturbation, while in the opposite case at $\tau < t_{\rm mer}$ the perturbation by a third black hole leads to an increase of the binary merger time. The angular momentum $j_p$ defining the boundary between these regimes is found from the solution of the equation $\tau(t,j_p) = t$. To transform the dependence on the semi-major axis to time and angular momentum $\tau(a) \rightarrow \tau(t,j)$, the binary merger time~\eqref{eq:tgw} are used. After the appropriate mathematical transformations we obtain
\begin{align} \label{eq:jp}
    j_p &= 141 \, Gm \left (\frac{\rho_{\eq} \delta_\pbh^4 t^{5/4}}{m \sigma c^{5/4}} \right)^{4/7} \\
    &\approx 0.28\,\delta_\pbh^{16/7} \left ( \frac{\sigma}{10\,\rm km \, s^{-1}} \right)^{-4/7} \left (\frac{m}{M_{\odot}} \right)^{3/7} \left (\frac{t}{\rm Gyr} \right)^{5/7}, \nonumber
\end{align}
where we used the gravitational focusing approximation in the cross section~\eqref{eq:cs_h}: the second term in the brackets of Eq.~\eqref{eq:tau} is the leading one. As can be seen from the comparison of the Eqs.~\eqref{eq:jch} and~\eqref{eq:jp}, most of the binaries that would merge in the contemporary Universe are perturbed since $j_p \gg j_{\rm ch}$. Thus, a constraint on the angular momentum of merging binaries $1 < j <j_{p}$ arises, as is clearly demonstrated by Fig.~\ref{fig:area_b}. Let us now define the moment of time $t_{\rm sup}$, starting from which the characteristic binaries are subject to perturbations, as a solution to the equation $\tau(a_{\rm ch}) = t$: 
\begin{equation} \label{eq:tsup}
    t_{\rm sup} \approx \frac{4.1 \, \rm Myr}{\delta_\pbh^{120/41}} \left ( \frac{\sigma}{10\,\rm km \, s^{-1}} \right)^{37/41} \left ( \frac{m}{M_{\odot}} \right)^{-19/41}.
\end{equation} 
This expression is also obtained in the gravitational focusing approximation; the geometric cross-section limit for completeness is provided in Table~\ref{tab:1}. Alternatively, this moment can be defined as the intersection of the $j_p$ and $j_{\rm min}$ curves, which is also reflected in Fig.~\ref{fig:area_b}. For $t < t_{\rm sup}$, the merger time $t_{\rm mer}$ turns out to be shorter than the perturbation time $\tau$ even for binaries with the largest possible semimajor axis. However, for $t > t_{\rm sup}$, the merger suppression mechanism comes into play due to gravitational perturbations of the binaries by the surrounding PBHs.

\section{Primordial black hole merger rate} \label{sec5}

The perturbations of PBH binaries lead to their dynamical decrease: soft binaries are destroyed, and hard binaries switch to an orbit with angular momentum $j\sim 1$, which increases their merger time to values much larger than the age of the Universe $t_0$. The central idea of this paper is that the perturbed binaries actually ``drop out'' from the merger process. This can be described as a continuous time evolution of the initial distribution~\eqref{eq:dp_dadj} in orbital parameters 
\begin{equation}
    dP = \frac{3}{4} \left ( \frac{\delta}{\overline{x}} \right)^{3/2} \frac{\sqrt{a}}{j^2} \, S(t,j,a) \, da dj,
\end{equation}
where $S$ is the suppression factor taking into account the decrease of the amount of merging binaries in time. Physically, it corresponds to the perturbation probability of the binary system and in our model is taken in the following form
\begin{equation} \label{eq:Sf}
    S(t,j,a) = 
    \begin{cases}
        \theta(j_{\rm cut} - j) e^{-t/\tau} + \theta(j - j_{\rm cut}) & \text{if } a \leq a_{\rm h}, \\
        e^{-t/\tau} & \text{if } a > a_{\rm h},
    \end{cases}
\end{equation}
where $\theta$ is Heaviside step function, $a_{\rm h}$ is the semimajor axis separating the hard and soft binaries defined by Eq.~\eqref{eq:ah} and $j_{\rm cut} = 1/2$ is the angular momentum above which the binaries are stable with respect to perturbations. The presented form of the suppression factor is physically explained in the previous section. The limitations of a sharp transition at angular momentum $j_p$ in the suppression model~\eqref{eq:Sf} are presented at the end of this section. A more physically realistic case, which must take into account the complex dynamics of binary-single PBH interaction at large binary angular momentum $j\sim1$, is discussed in the appendix A. Note that the influence of the initial clustering of PBHs on their merger rate was also considered in Refs.~\cite{Bringmann:2018mxj, Young:2019gfc,Atal:2020igj,Crescimbeni:2025ywm}, but without taking into account the perturbations of the binaries. We also note recent paper~\cite{Holst:2024ubt} that studied the PBH mergers in clusters during the era of PBH domination in the very early Universe. 

\begin{figure}%[h]
	\begin{center}
\includegraphics[angle=0,width=0.48\textwidth]{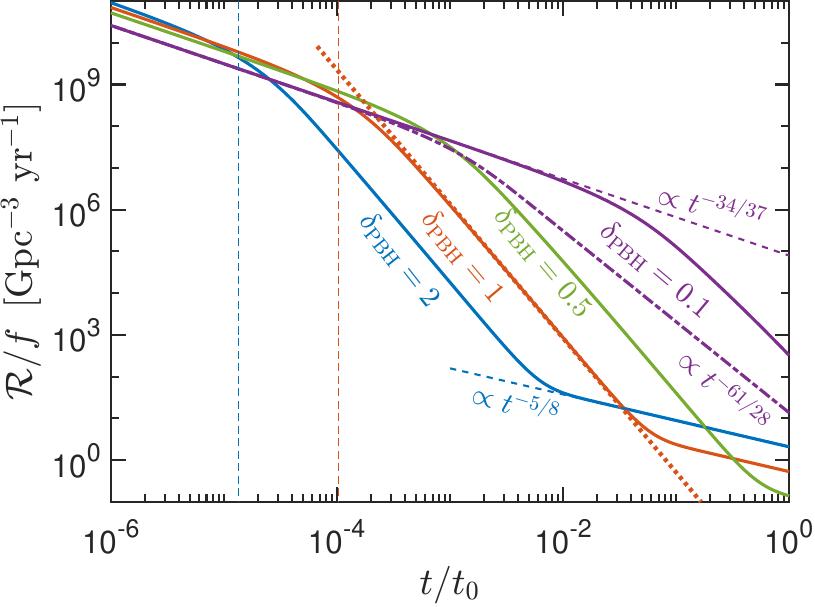}
	\end{center}
\caption{Time evolution of the merger rate for different values of the density contrast $\delta_\pbh$ and velocity dispersion in the cluster $\sigma = 10$~km~s$^{-1}$ is shown by the solid lines. The purple dot-dashed line shows the case of a cluster with $\sigma = 10^3$~km~s$^{-1}$ and $\delta_\pbh = 0.1$. The red dotted line shows the analytical estimate of the merger rate given by Eq.~\eqref{eq:mr_est}. The dashed purple line shows the PBH merger rate when the effects of binary perturbations are negligible (corresponding to the unclustered case). The vertical dashed lines show the moment of time $t_{\rm sup}$, determined by Eq.~\eqref{eq:tsup}, for $\delta_\pbh = 2$ and $\delta_\pbh = 1$, respectively.}
	\label{gr:mr} 
\end{figure}

To determine the merger rate, it is convenient to transform from the semimajor axis variable $a$ to time $t$ using Eq.~\eqref{eq:tgw}. Then the differential probability of a merger per unit time is given by 
\begin{equation}\label{eq:dPdt_Sf}
    \frac{dP}{dt} = \frac{3}{16} \frac{\delta_\pbh^2}{T} \left ( \frac{t}{T} \right)^{-5/8} \int \limits_{j_{\rm min}}^{1} j^{-37/8} S(t,j) \, dj,
\end{equation}
where the lower limit is defined as $j_{\rm min} = \left ( t/T \right)^{3/37} \delta_\pbh^{16/37}$ (which is also equivalent to Eq.~\eqref{eq:jch}), and
\begin{equation}
    T = \frac{3 c^5 \overline{x}^4 \delta_\pbh^{4/3}}{170 G^3 m^3} \sim 10^{34} \left ( \frac{m}{M_{\odot}} \right)^{-5/3} \, \rm{yr},
\end{equation}
physically corresponds to the merger time of a PBH binary in a circular orbit with the maximum possible semi-major axis in the case of $\delta_\pbh = 1$. The comoving merger rate of PBHs is determined by the expression
\begin{equation}\label{eq:mr}
    \mathcal{R} = \frac{f \Omega_{\rm DM}}{m} \frac{3 H_0^2}{8 \pi G}  \frac{dP}{dt},
\end{equation}
where the second fraction is the critical density $\rho_{\rm crit}$ and after numerical integration of Eq.~\eqref{eq:dPdt_Sf} we obtain the time dependence of the merger rate, which is shown in Fig.~\eqref{gr:mr}. Note that we normalized all results here by $f$ to exclude the dependence on the fraction of PBHs in dark matter. One can see that the merger rate at $t \ll t_0$ follows the dependence known for the case of unclustered PBHs~\cite{Raidal:2018bbj}\footnote{Although we note that the effects of natural Poisson clustering of PBHs at the matter dominated stage slightly modify this dependence, but they are valid at the fraction of PBHs in the dark matter composition $f \gtrsim 0.01$~\cite{Stasenko:2024pzd}}, the explanation is that the binaries have not yet begun to perturb, i.e. $S = 1$ and integration of the power function in Eq.~\eqref{eq:dPdt_Sf} easily gives this answer. The merger rate then drops sharply at $t > t_{\rm sup}$, with this behavior depending on the velocity dispersion in the cluster, as can be seen by comparing the purple dashed line and the solid lines. 

\begin{figure}%[h]
	\begin{center}
\includegraphics[angle=0,width=0.48\textwidth]{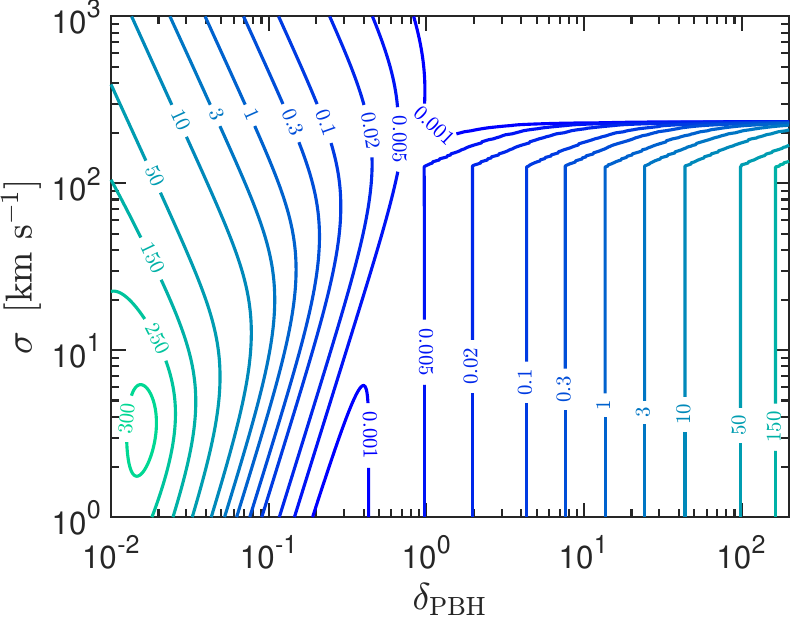}
	\end{center}
\caption{Contour plot of the current merger rate of PBH binaries $\mathcal{R}_0$ (in units of [Gpc$^{ -3}$~yr$^{-1}$]) calculated for the PBH fraction $f = 0.01$ for different clustering parameters $\delta_\pbh$ and the velocity dispersion $\sigma$ in the cluster, the PBH mass $m = 10\,M_{\odot}$. Regions in parameter space incompatible with the LIGO-Virgo-KAGRA merger rate $\mathcal{R}_0 = 18 - 44 $~Gpc$^{ -3}$~yr$^{-1}$~\cite{KAGRA:2021duu} are ruled out by GW observations. For other PBH fractions $f \neq 0.01$, the merger rate should be rescaled by a factor of $f/0.01$ (also the condition $f \leq \delta_\pbh$ should be satisfied).}
	\label{gr:mrdv} 
\end{figure}

To analyze the obtained result, we derive an approximate expression for the merger rate in the time region where the suppression factor is of the form $S = e^{-t/\tau(j,t)}$. Also for the beginning we will consider that the velocity dispersion of PBHs in the cluster is insignificant, i.e. the second term of gravitational focusing in the cross section~\eqref{eq:tau} is the leading one. As can be seen from Fig.~\eqref{fig:area_b}, the main contribution to the integral~\eqref{eq:dPdt_Sf} comes from the binaries with angular momenta $j \gtrsim j_p$. Numerical analysis shows that a good approximation is achieved by truncating the integral~\eqref{eq:dPdt_Sf} from below at the value $j = 0.82\, j_p$, and the suppression factor in the integration region can be neglected $e^{-t/\tau(j,t)} \rightarrow 1$. After the corresponding calculations we obtain
\begin{align} \label{eq:mr_est}
    \mathcal{R} &\approx \frac{1.8 \times 10^6}{\rm yr\, Gpc^3} f \delta_\pbh^{-44/7} \left ( \frac{t}{10^8 \, \rm yr} \right)^{-45/14} \nonumber \\
    &\times \left ( \frac{m}{M_{\odot}} \right)^{-27/14} \left ( \frac{\sigma}{10 \,\rm km \, s^{-1}} \right)^{29/14}.
\end{align}
This dependence is shown by the red dotted line in Fig.~\eqref{gr:mr}, from which we can see that the derived expression indeed approximates well the merger rate at $t > t_{\rm sup}$. Let us now consider the limiting case of a cluster with huge velocity dispersion. In this case, the cross section~\eqref{eq:cs_h} becomes the geometric $\Sigma \propto a^2$. Using Eqs.~\eqref{eq:tgw} and~\eqref{eq:tau} we obtain that the characteristic time of the perturbations scales as $\tau \propto t^{-1/2} j^{7/2}$. As in Sec.~\ref{sec4}, determining the limiting angular momentum $\tau(t,j_p) = t$ leads to the following relation $j_p \propto t^{3/7}$. Truncating the integral~\eqref{eq:dPdt_Sf} from below at angular momentum $\sim j_p$ (similar to the derivation Eq.~\eqref{eq:mr_est}) gives the time dependence of the merger rate $\mathcal{R} \propto t^{-61/28}$. Note that in this case the power-law decay of the merger rate will change to exponential over time when the condition $j_p(t) = 1$ is reached, as follows from Eq.~\eqref{eq:Sf}. This is due to the fact that binaries with a sufficiently long lifetime are soft in the cluster. That is, even PBH binaries in circular orbits (and even more so in elliptical orbits) will be destroyed due to perturbations, which is also seen in Fig.~\ref{gr:mrdv} in the region of $\delta_\pbh > 1$ and $\sigma > 100$~km~s$^{-1}$

Let us now turn to the discussion of larger time scales, where the merger rate experiences a sharp change in slope and comes to a dependence $\mathcal{R} \propto t^{-5/8}$. This result is more of an artifact of our model rather than a physical effect. This behavior arises from the specific form of the suppression factor~\eqref{eq:Sf}, where we assume that the orbital parameters of binaries with sufficiently large angular momenta $j > j_{\rm cut}$ do not change under perturbations. Then it is this region of angular momenta that will give the dominant contribution to the integral~\eqref{eq:dPdt_Sf}, so it is evaluated by a dimensionless constant. The time dependence arises only from the pre-integral factor, which explains the result. Moreover, changing the value of $j_{\rm cut}$ will only affect the amplitude of the merger rate, not its logarithmic slope. As noted in Section~\ref{sec4}, a correct description requires taking into account the change in the distribution of the orbital parameters of the binaries under perturbations. In Appendix A, we develop a simple physical model for the merger of these tight PBH binaries, taking into account the evolution of the distribution of their parameters $a$ and $j$ due to perturbations in clusters. In particular, assuming that the angular momentum distribution becomes the power-law form $dP/dj \propto j^{n}$, the merger rate will vary as $\mathcal{R} \propto t^{(n - 6)/7}$ (see also Ref.~\cite{Franciolini:2022ewd}), with $n = 1$ corresponding to the thermal distribution. Thus, in the physically realistic case, we expect the merger rate to also vary only slightly with cosmological time. However, the main result of this work is that there is a time window in which there is a sharp drop in the merger rate, which significantly diverges from the predictions of models in which the PBHs are initially distributed according to a Poisson law in space. This provides an opportunity to test the clustering of PBHs in future GW observations. 

As noted above, perturbations of binaries with small eccentricities (almost circular orbits) make them harder, hence the merger time will decrease and therefore the merger rate of such binaries will increase. Therefore, our results in Fig.\ref{gr:mr} for $\delta_\pbh > 1$ and time scales $\sim t_0$ (in this case, binaries with $j > j_{\rm cut}$ merge) should be considered as a conservative lower estimate. Figure ~\ref{gr:mrdv} shows the contour plot of the current merger rate as a function of the clustering parameter $\delta_{\rm PBH}$ and the velocity dispersion $\sigma$ in the cluster. For the case $f \simeq 0.01$ the majority of the clustering parameter space is consistent with current GW observation. Also one can see that there exists a region of parameters near $\delta_\pbh \gtrsim 1$, where one can explain all dark matter in terms of PBHs with mass $\sim 10\,M_{\odot}$. However, many other constraints imposed on the PBHs of these masses exclude this possibility. With a significant increase in the clustering parameter $\delta_\pbh \gg 1$, the merger rate is determined by binaries with nearly circular orbits. Moreover, increasing $\delta_\pbh$ leads to an enhancement of the current merger rate, as can also be seen in Fig.~\ref{gr:mr}. Thus, the early clustering effect makes it virtually impossible to relax the GW constraints to the level where all dark matter could consist of PBHs. This conclusion is also consistent with the results obtained in Ref.~\cite{Bringmann:2018mxj}. However it should be noted, that in the region $\delta_\pbh \gtrsim 1$ and $\sigma \gtrsim 100$~km/s, the merger rate is exponentially suppressed (according to Eq.~\eqref{eq:Sf}), since in this case all binary systems are soft and undergo destruction. In the opposite limit $\delta_\pbh \ll 1$ clusters are formed at enough late stages, that significantly reduces the efficiency of perturbations on binary systems (see Eq.~\eqref{eq:tsup}), and we return to the situation of PBHs unclustered at birth, for which the standard constraints on their fraction in dark matter $f$ are valid.

\section{Stochastic gravitational wave background}\label{sec6}

In the previous section the behavior of the PBH merger rate on cosmological time scales was considered. However individual events of black hole mergers at high redshifts may remain observationally unresolved, nevertheless their total contribution to the energy density of GW in the Universe can be significant. In this section, we discuss the impact of our results on the characteristics of the GW background, which is traditionally described by the relative spectral energy density~\cite{Phinney:2001di}
\begin{align}\label{eq:Ogw}
    \Omega_{\rm gw} &= \frac{1}{\rho_{\rm crit} c^2} \frac{d \rho_{\rm gw}}{d \ln \nu} \nonumber \\
    &= \frac{1}{\rho_{\rm crit} c^2} \int_0^{\infty} dz \, \frac{\mathcal{R}(z)}{H(z) (1 + z)^2} \frac{d E_{\rm gw}}{d \ln \nu_s},
\end{align}
where $\nu_s = (1 + z) \nu$ is the GW frequency in the source system and $dE_{\rm gw}/d \ln \nu_s$ is the GW energy emitted during the merger of a binary black hole in a logarithmic bin of frequency. In this paper we use the inspiral-merger-ringdown energy spectrum~\cite{Ajith:2007kx, Ajith:2009bn}. In fact, Eq.~\eqref{eq:Ogw} is the sum of the energy released from all merger events summed over all redshifts. It should be noted that the analysis of the stochastic GW background from PBH mergers has also been studied in Refs.~\cite{Mandic:2016lcn,Wang:2016ana,Clesse:2016ajp, Atal:2020igj,Bavera:2021wmw,Mukherjee:2021ags,Garcia-Bellido:2021jlq,Braglia:2021wwa,Atal:2022zux}.

First, we parametrize the redshift evolution of the merger rate as $\mathcal{R} = \mathcal{R}_0 (1 + z)^{\kappa}$ and derive an analytic expression for the GW background, which will show the main idea of this section. For these purposes it is enough for us to consider only the inspiral phase of a binary black hole whose energy spectrum is~\cite{Maggiore:2007ulw} 
\begin{equation} \label{eq:dEgwdlnf}
    \frac{d E}{d \ln \nu_s} = \frac{\pi^{2/3}}{3 G} (G M_c)^{5/3} \nu_s^{2/3},
\end{equation}
where $M_c = (m_1 m_2)^{3/5}/(m_1 + m_2)^{1/5}$ is the chirp mass and in our case of equal masses black holes $M_c = m / 2^{1/5}$. We will assume that the inspiral phase, described by Eq.~\eqref{eq:dEgwdlnf}, continues up to some maximum frequency $\nu_{\rm max}$ and then $dE / d\ln \nu_s$ drops to zero. Such simplified model corresponds to the instantaneous formation of a new merged black hole after reaching $\nu_{\rm max}$. Then the amplitude of the GW background depends on the frequency as follows 
\begin{align} \label{eq:Ogw2}
    &\Omega_{\rm gw} = A \nu^{2/3} \frac{\mathcal{R}_0 \pi^{2/3} (G m)^{5/3}}{3 \sqrt{2} G \rho_{\rm crit} c^2} \int \limits_{0}^{(\nu_{\rm max}/\nu)-1} \frac{(1 + z)^{\kappa +2/3}}{(1+z)^2H(z)}  \, dz \nonumber \\
    &\approx \frac{A \mathcal{R}_0 \pi^{2/3} (G m)^{5/3} \nu^{2/3}}{3 \sqrt{2} G \rho_{\rm crit} c^2 H_0 \sqrt{\Omega_M} (\kappa -11/6)}  \left [ \left ( \frac{\nu_{\rm max}}{\nu}  \right)^{\kappa - 11/6} -1  \right],
\end{align}
where in the second equality we used the Hubble rate approximation $H(z) \approx H_0 \sqrt{\Omega_M (1 + z)^3}$, which holds already at redshifts $z>1$. The introduced constant $A$ compensates for the deviations of the simplified spectrum~\eqref{eq:dEgwdlnf} from the realistic case and will be determined numerically in the further analysis. It can be seen that at $\kappa < 11/6$ the second term in brackets of Eq.~\eqref{eq:Ogw2} is the dominant one, which leads to the classical power-law spectrum of $\Omega_{\rm gw} \propto \nu^{2/3}$, which is characteristic for black hole mergers. However, when $\kappa > 11/6$, the spectral slope deviates from the canonical scaling, taking the form $\Omega_{\rm gw} \propto \nu^{5/2 - \kappa}$. Note that earlier work~\cite{Atal:2022zux} also showed a direct connection between the spectral index of the GW background and the redshift evolution of black holes merger rate. Of particular interest is the case of sufficiently fast growth of the merger rate with redshift $\kappa > 5/2$, when the spectral slope of the GW energy density becomes negative --- this is the case realized in this work.

\begin{figure}%[h]
	\begin{center}
\includegraphics[angle=0,width=0.48\textwidth]{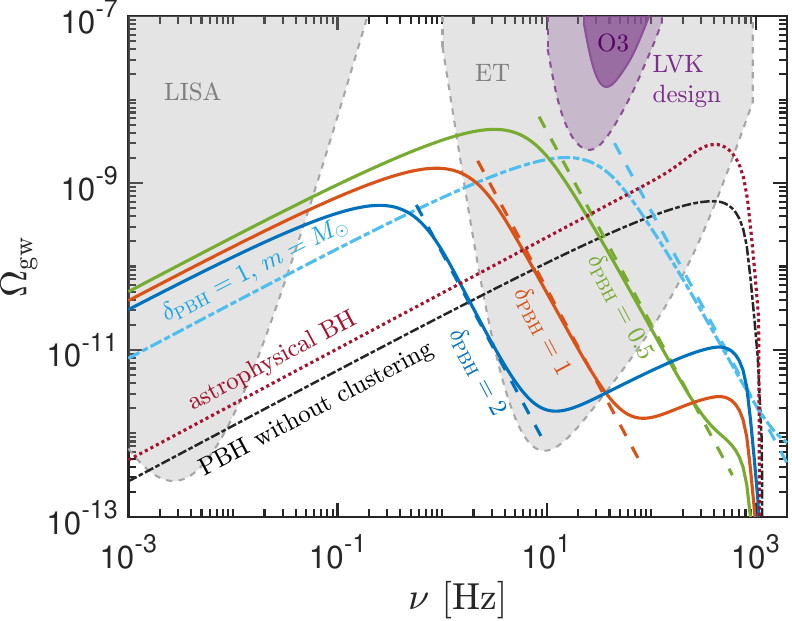}
	\end{center}
\caption{Spectral energy density of the stochastic GW background from mergers of PBHs with $m = 10\,M_{\odot}$ mass for different values of the clustering parameter $\delta_\pbh$ and $\sigma = 10$~km~s$^{-1}$. For illustration purposes we adopt $f = 0.1$ here, though  microlensing constraints typically favor $f \sim 0.01$. The dashed line shows the analytic approximation is given by Eq.~\eqref{eq:Ogw3}, with the minimum plotting frequency determined by $\nu_{\rm sup}$ from Eq.~\eqref{eq:nu_sup}. The dash-dotted line line shows the case of the Poisson (with no initial clustering) distribution of PBHs normalized such that the current merger rate $\mathcal{R}_{\rm PBH}(t_0) = 10$~Gpc$^{-3}$~yr$^{-1}$ and changes with redshift as $\mathcal{R} \propto t^{-34/37}$. The red dotted line shows the case of astrophysical black holes (see Appendix B for details) and $\mathcal{R}_{\rm ABH}(t_0) = 40$~Gpc$^{-3}$~yr$^{-1}$. The shaded areas show the projected sensitivity of some future GW detectors, as well as the upper limit from the O3 run results~\cite{KAGRA:2021kbb}.}
	\label{gr:gwb} 
\end{figure}

Figure~\ref{gr:gwb} shows the stochastic GW background calculated using the merger rate~\eqref{eq:mr} for different values of the clustering parameter $\delta_\pbh$. As can be seen from the graph, the spectral index takes negative values in a certain frequency range. We now obtain an analytical approximation for $\Omega_{\rm gw}$ in this region. As shown in the previous section, during the period of active perturbation of PBH binaries the merger rate evolves according to Eq.~\eqref{eq:mr_est} as $\mathcal{R} \propto (t/t_0)^{-45/14}$, which corresponds to the following dependence on the redshift $z$
\begin{equation}
    \mathcal{R} = \mathcal{R}_0 \left ( \frac{2}{3H_0 t_0 \sqrt{\Omega_M}} \right)^{-45/14} (1 + z)^{135/28},
\end{equation}
where, as before, we have used the dust stage approximation $t(z) = 2/3H(z)$, since we are interested in the high redshift limit. Taking the maximum frequency of the GW in Eq.~\eqref{eq:Ogw2} as $\nu_{\rm max} = 0.057 \, c^3/Gm$ (corresponding to the spectrum break $dE_{\rm gw}/d \ln{\nu_s}$ at higher frequencies~\cite{Ajith:2007kx}), that approximation means that the inspiral phase is formally extended to the ringdown frequency. Good agreement with numerical calculations is then obtained with the estimate $A \sim 1$:
\begin{align} \label{eq:Ogw3}
    \Omega_{\rm gw} &\approx 9.7 \times10^{-7} f \delta_\pbh^{-44/7} \left ( \frac{\sigma}{10 \,\rm km \, s^{-1}} \right)^{29/14} \nonumber \\
    &\times \left ( \frac{m}{M_{\odot}} \right)^{-13/4} \left ( \frac{\nu}{10 \, \rm{Hz}} \right)^{-65/28}.
\end{align}
This dependence is shown by dotted lines in Fig.~\ref{gr:gwb}. The suppression of the merger rate begins at time $t_{\rm sup}$, defined by Eq.~\eqref{eq:tsup}, so the corresponding GW frequency experiences a redshift $\nu_{\rm sup} = \nu_{\rm max}/(1+z_{\rm sup})$:
\begin{align}\label{eq:nu_sup}
    \nu_{\rm sup} &= \nu_{\rm max} \left ( \frac{3H_0t_{\rm sup} \sqrt{\Omega_M}}{2} \right)^{2/3}  \\
    &\approx 2.2 \, \delta_\pbh^{-80/41} \left ( \frac{\sigma}{10 \,\rm km \, s^{-1}} \right)^{74/123} \left ( \frac{m}{M_{\odot}} \right)^{-161/123} \, \rm{Hz}. \nonumber
\end{align}
Thus, in the frequency range $\nu > \nu_{\rm sup}$, the spectral density of GW is characterized by a negative slope $\Omega_{\rm gw} \propto \nu^{-65/28}$. At lower frequencies, the classical dependence $\Omega_{\rm gw} \propto \nu^{2/3}$, typical for merging black hole binaries. However, in the frequency band of space-based interferometers like LISA, a excess of $\Omega_{\rm gw}$ is expected compared to the case of PBHs with Poisson spatial distribution and even astrophysical black hole mergers are shown by the red dotted line in Fig.~\ref{gr:gwb}. Thus, the detection of strong GW background in this frequency range, coupled with the absence of a corresponding signal on ground-based detectors (when interpolated to higher frequencies $\Omega_{\rm gw} \propto \nu^{2/3}$) will hint at the presence of PBHs, which must be highly clustered. However note that if the PBH fraction $f \ll 0.1$, such features in the GW energy spectrum will be unobservable against other cosmological GW contributions. Let us now discuss the frequencies $\sim 100$~Hz. Provided that PBHs of tens solar masses do not dominate the observed GW events, then at these frequencies the main contribution to the GW background comes from mergers of black holes of stellar origin~\cite{Mandic:2016lcn}, which also leads to the standard dependence $\Omega_{\rm gw} \propto \nu^{2/3}$, as can be seen in Fig.~\ref{gr:gwb}. However, a comprehensive analysis of the GW background taking into account various PBHs clustering scenarios and the contribution of astrophysical sources is beyond the scope of this paper and will be addressed in a separate article. Nevertheless, as can be seen from Fig.~\ref{gr:gwb}, there is parameter region in which future GW detectors will be able to test the possibility of PBHs clustering.

\begin{table*}
\caption{
Scaling of the merger rate $\mathcal{R}$ and stochastic GW background $\Omega_{\rm gw}$ for different cases of black hole mergers: with and without initial clustering for PBHs, and astrophysical black holes. For the clustered PBH scenario, the scaling for $\mathcal{R}$ and $\Omega_{\rm gw}$ applies after the suppression time $t_{\rm sup}$ and frequency $\nu_{\rm sup}$, respectively (see Eqs.~\eqref{eq:tsup} and~\eqref{eq:nu_sup}). Here, we introduce the notation  $\sigma_{10}$ is the velocity dispersion in units of $10$~km~s$^{-1}$ and $m_1$ is the PBH mass in $M_{\odot}$, in order to write the expression for $t_{\rm sup}$. 
\label{tab:phases}}
\begin{ruledtabular}
\begin{tabular}{l c c c c }

& PBH clustering with low $\sigma$ & PBH clustering with high $\sigma$ & PBH without clustering & ABH\\
\hline
$\mathcal{R}$ & $\propto t^{-45/14}$ & $\propto t^{-61/28}$ & $\propto t^{-34/37}$ & see Fig.~\ref{gr:abhpbh_mr} \\
$\Omega_{\rm gw}$  & $\propto \nu^{-65/28}$ & $\propto \nu^{-43/56}$ & $ \propto \nu^{2/3}$ & $ \propto \nu^{2/3}$\\
$t_{\rm sup}$ [Myr] & $4.1 \, \delta_\pbh^{-120/41} \sigma_{10}^{37/41} m_1^{-19/41}$ & $5.9 \, \delta_\pbh^{-92/45} \sigma_{10}^{-37/45} m_1^{-1/45}$  & --- & --- \\
$\nu_{\rm sup}$ [Hz] & $2.2 \, \delta_\pbh^{-80/41} \sigma_{10}^{74/123} m_1^{-161/123}$ & $2.8 \, \delta_\pbh^{-184/135} \sigma_{10}^{-74/135} m_1^{-2/135}$  & --- & --- \\
\end{tabular}
\end{ruledtabular}
\label{tab:1}
\end{table*}

\section{Conclusion}\label{sec7}

The paper investigated the influence of the initial clustering of PBHs on the dynamics of their mergers. It was shown that a significant part of the binaries are characterized by sufficiently wide orbits and small angular momenta, and therefore they are perturbed in the dense environment of the cluster. This process leads to an effective ``elimination'' of such binaries from the merger process due to a significant increase of their coalescence time or even destruction. At the same time, rare close binaries are more likely to avoid perturbations and therefore continue to merge. Such a change in the distribution of the PBH binaries orbital parameters causes a suppression of the merger rate starting from the time $t_{\rm sup}$ determined by Eq.~\eqref{eq:tsup}, these results are shown in Fig.~\ref{gr:mr}. In the epoch of active binaries perturbation, the merger rate scales with cosmological time as $\mathcal{R} \propto t^{-45/14}$ (in accordance with Eq.~\eqref{eq:mr_est}). Which differs significantly from the case of a Poisson (initially non-clustered) distribution of PBHs in the space $\mathcal{R} \propto t^{-34/37}$. The study of the redshift evolution of black holes merger rate is one of the target for the next generation of GW detectors. Therefore, future observations will be able not only to verify the existence of PBHs, but also to reconstruct the features of their initial spatial distribution, which will make it possible to identify the most plausible scenarios for their formation. The absence of GW signals at high redshifts, in turn, will impose strict constraints on the parameters of PBHs formation models.

On the other hand, if direct detection of GW transients from black hole mergers at large redshifts turns out to be observationally difficult, their overall contribution may still be significant as the integral effect on the GW energy in the Universe. Therefore, in Section~\ref{sec6} we analyze the impact of our results for the stochastic GW background. We demonstrate that a rapid growth in the black hole merger rate with redshift causes spectral features in the GW energy density, as illustrated in Fig.~\ref{gr:gwb}. Specifically, when the merger rate is scaled with time as $\mathcal{R} \propto t^{-45/14}$, the stochastic GW background depends on the frequency $\Omega_{\rm gw} \propto \nu^{-65/28}$ in the frequency range above $\nu_{\rm sup}$ given by Eq.~\eqref{eq:nu_sup}. This is radically different from the traditional expectation of $\Omega_{\rm gw} \propto \nu^{2/3}$ for black hole mergers. We summarize our main findings relating to the scaling of the merger rate and the GW background in Table~\ref{tab:1}. The obtained results provide an additional possibility for experimentally testing PBH clustering scenarios. 

\section*{Acknowledgment}
The work was funded by the Ministry of Science and Higher Education of the Russian Federation, Project ``New Phenomena in Particle Physics and the Early Universe'' FSWU-2023-0073. I am grateful to the anonymous reviewers for their constructive comments and recommendations, which helped improve the paper. 

\section*{Appendix A: Mergers of perturbed binaries}

To ensure completeness in Sec.~\ref{sec5}, this appendix outlines considerations relating to the merging of perturbed binaries. First of all, it is necessary to separate hard binaries from soft ones in the entire population of PBH binaries. For this purpose, it is sufficient to consider the distribution of distances between two PBHs $dP = 3x^2/\overline{x}^3\, dx$ and integrate it to $x(a_h)$:
\begin{align}
     &P_h \equiv P(a < a_h)  = \left ( \frac{a_h \delta_\pbh}{\overline{x}} \right)^{3/4}  \nonumber \\
     & \simeq 2 \times 10^{-3} \delta_\pbh \left ( \frac{m}{M_{\odot}} \right)^{1/2} \left ( \frac{\sigma}{10 \, \rm km \, s^{-1}} \right)^{-3/2}.
\end{align}
The relationship between $x$ and the major semiaxis $a$ is determined by Eq.~\eqref{eq:axby}, and $a_h$ is the major semiaxis separating hard binaries from soft ones~\eqref{eq:ah}.  

\begin{figure}%[h]
	\begin{center}
\includegraphics[angle=0,width=0.48\textwidth]{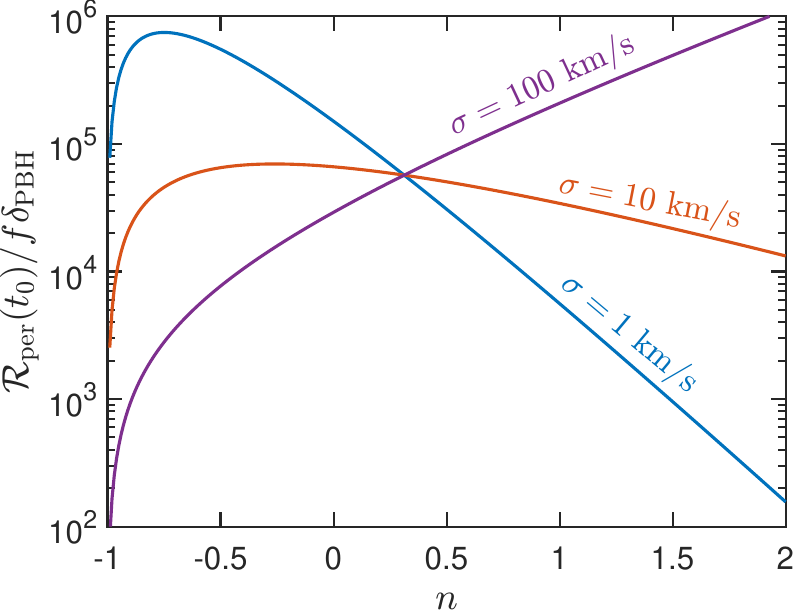}
	\end{center}
\caption{The current merger rate as a function of the angular momentum distribution parameter $n$ in Eq.~\eqref{dP_per}. Different curves correspond to different velocity dispersions $\sigma$.}
	\label{gr:mr_per} 
\end{figure}

Now, following the explanation in Ref.~\cite{Heggie_Hut_2003}, we will take a qualitative look at the dynamics of hard binaries in clusters. The change in the energy of a hard binary resulting from a single scattering event involving a PBH is comparable in magnitude to its binding $\Delta E \sim |\varepsilon| = Gm^2/2a$. The conservation of energy during three-body scattering in the rest frame of the barycenter system reads
\begin{equation}
	\frac{\mu v^2}{2} + \varepsilon = \frac{\mu v'^2}{2} + \varepsilon (1 + \Delta),
\end{equation}
here, $v$ and $v'$ are the relative velocities at infinity of the binary and single PBH, respectively, before and after scattering, $\mu = m/3$ is the reduced mass and $\Delta = (\varepsilon'/\varepsilon - 1)\simeq 1$ is the relative change in binding energy of binary after scattering. In the case of a sufficiently hard binary, i.e. when $|\varepsilon| \gg \mu v^2$, the relative velocity after scattering is given by $v'\simeq \sqrt{-3 \, \varepsilon\Delta/m}$ and the velocity of the binary will be $v'_b = v'/3$. To obtain the change in velocity relative to the whole cluster, it is necessary to add the velocity of the triple system, but this is negligible within the limits of a sufficiently hard binary. Thus, the fate of the binary system is as follows: due to three-body scattering, the binary becomes increasingly bound until it ultimately acquires sufficient recoil to leave the cluster. Also, since the recoil velocity depends solely on the binding energy, we assume as a first approximation that the perturbed binaries have approximately the same binding energy $\varepsilon = km\sigma^2/2$ and merge outside the cluster. Moreover, due to perturbations, their initial angular momentum distribution is strongly mixed (and likely approaches to thermal one), so we assume the distribution of binary parameters to be as follows:
\begin{equation}\label{dP_per}
    dP_{\rm per} = P_h (n+1) \delta(a - a_*) j^n \,da dj,
\end{equation}
where $a_* = Gm/k\sigma^2$, $\delta(a - a_*)$ is the delta function and $k \gg 1$ (for globular star clusters $k\simeq 100$ typically~\cite{Heggie_Hut_2003, 2008gady.book.....B}). Note that the case $n = 1$ corresponds to thermal distribution. The fraction of binaries merging by time $t$ is defined as $P_{\rm per} = \int_0^{j_*} dP_{\rm per} = P_h j_*^{n+1}$, where
\begin{equation}
	j_* = \left ( \frac{170 \, G^3 m^3 t}{3 \, c^5 a_*^4} \right)^{1/7}
\end{equation}
is obtained using the merger time~\eqref{eq:tgw}. Here, for simplicity, we consider the case that perturbed binaries formed in a enough young Universe. In other words, the time difference between the formation epoch and the merger of these binaries corresponds to cosmological time. However, it is clear that this assumption may be inaccurate, particularly for binaries that initially formed a very close system (their characteristic perturbation time is large). The merger rate is determined by a similar expression to that given in Eq.~\eqref{eq:mr}
\begin{align}\label{eq:mr_p1}
    &\mathcal{R}_{\rm per} = \frac{f \Omega_{\rm DM}}{m} \frac{3 H_0^2}{8 \pi G}  \frac{dP_{\rm per}}{dt} \nonumber  \\
    &= \frac{\rho_{\rm DM} f \delta_\pbh}{m} \frac{n+1}{7t} \left ( \frac{Gm^{2/3} \rho_\eq}{2 \sigma^2} \right)^{3/4} \left ( \frac{170 k^4\sigma^8 t}{3c^5 Gm} \right)^{(n+1)/7}. 
\end{align}
For this result to be consistent, a transition to the unperturbed binary merger regime, defined by Eq.~\eqref{eq:mr_est}, must occur, but this analysis is beyond the scope of this paper. By numerically estimating the merger rate~\eqref{eq:mr_p1} for some parameters, we obtain
\begin{widetext}
\begin{align} \label{eq:mr_p2}
    \mathcal{R}_{\rm per} &\simeq  \dfrac{3.4 \times 10^4}{\rm Gpc^3 yr} f \delta_\pbh \left ( \dfrac{k}{100} \right)^{8/7} \left ( \dfrac{m}{10\,M_{\odot}} \right)^{-11/14} \left ( \dfrac{\sigma}{10\, \rm km\,s^{-1}} \right)^{11/14}  \left ( \dfrac{t}{t_0} \right)^{-5/7} \quad {\rm{for}} \, n= 1, \nonumber \\
    \mathcal{R}_{\rm per} &\simeq \dfrac{6.6 \times 10^4}{\rm Gpc^3 yr} f \delta_\pbh \left ( \dfrac{k}{100} \right)^{4/7} \left ( \dfrac{m}{10\,M_{\odot}} \right)^{-9/14} \left ( \dfrac{\sigma}{10\, \rm km\,s^{-1}} \right)^{-5/14}  \left ( \dfrac{t}{t_0} \right)^{-6/7} \quad {\rm{for}} \, n= 0,
\end{align}
\end{widetext}
where we consider two illustrative examples for angular momentum distribution. However, we note that depending on the velocity dispersion $\sigma$, we note that the the merger rate can vary by several orders of magnitude as a function of $n$, as shown in Fig.~\ref{gr:mr_per}. We also note that, for all reasonable values of the parameter $n$ in Eq.~\eqref{dP_per}, the merger rate of perturbed binaries is a slow function of cosmological time. Therefore, their merging does not lead to the appearance of spectral features in the stochastic GW background, similar to those discussed in Sec.~\ref{sec6}.

\section*{Appendix B: Mergers of stellar origin black holes} 

\begin{figure}%[h]
	\begin{center}
\includegraphics[angle=0,width=0.48\textwidth]{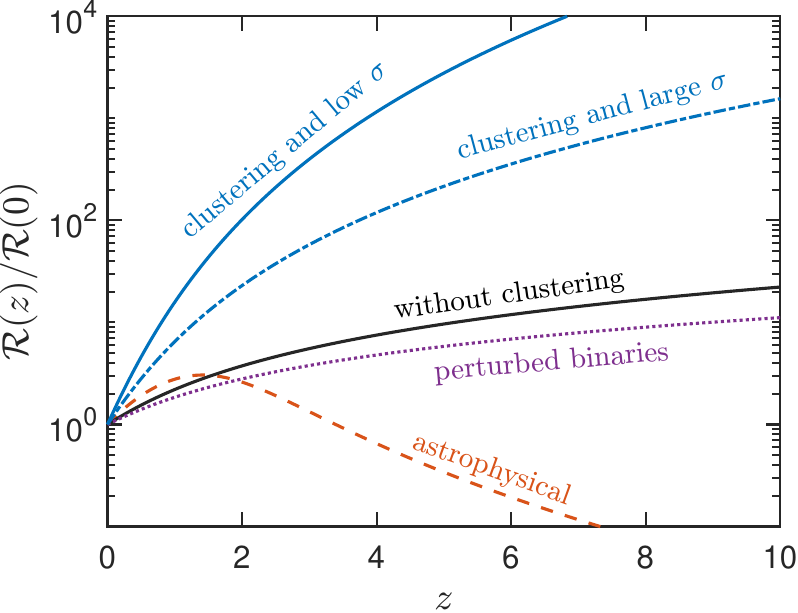}
	\end{center}
\caption{The normalized merger rate of binary black holes of different origins. The red dashed line corresponds to astrophysical black holes defined by Eq.~\eqref{eq:mrabh}. The blue solid and dash-dotted lines show the mergers of unperturbed PBHs in the case of their initial clustering, which are scaled as $\mathcal{R} \propto t^{-45/14}$ and $\mathcal{R} \propto t^{-61/28}$, respectively (see Sec.~\ref{sec5} for details). The dotted purple line represents the merging of perturbed PBH binaries with the thermal distribution of angular momentum $\mathcal{R} \propto t^{-5/7}$ ($n=1$ in Eqs.~\eqref{dP_per}~and~\eqref{eq:mr_p1}). The black solid line shows mergers of PBHs without any initial clustering in their spatial distribution $\mathcal{R}\propto t^{-34/37}$.}
	\label{gr:abhpbh_mr} 
\end{figure}

This appendix considers a more traditional source of GW events: mergers of astrophysical black holes. Here, we will follow the analysis of~\cite{Raidal:2018bbj,Mukherjee:2021ags}. The astrophysical black hole merger rate should follow the rate of star formation, which is fitted by the empirical formula~\cite{Madau:2014bja} 
\begin{equation}\label{eq:sfr}
	\mathcal{R}_{\rm{SFR}}(z) \propto \frac{(1+z)^{2.7}}{1 + \big ((1 + z)/2.9 \big)^{5.6}}. 
\end{equation}
Since the formation of stellar binaries, the subsequent formation of black holes, and their mergers are separated in time, a time delay must be introduced between these processes. Similar to Ref.~\cite{Belczynski:2016obo}, we assume that the delay time distribution is $P_{\rm del} \propto t^{-1}$ and zero otherwise. Then the dependence of the merger rate of astrophysical black holes on redshift is determined by 
\begin{align}\label{eq:mrabh}
	&\mathcal{R}_{\rm{ABH}}(z) \propto \int_z^{\infty} dz_f \, \frac{dt_f}{dz_f} \, dt_{\rm del} \, \mathcal{R}_{\rm{SFR}}(z_f) \nonumber \\
    &\times P_{\rm del}(t_{\rm del}) \delta \Big (t(z) - t(z_f) - t_{\rm del} \Big),
\end{align}
where $z_f$ is the redshift of the formation of the stellar binary and $dt/dz = (H(z)(1+z))^{-1}$. 

Figure~\ref{gr:abhpbh_mr} shows the temporal evolution of the merger rate of astrophysical and primordial black holes, taking various clustering scenarios into account. As can be seen, mergers of PBHs inevitably dominate astrophysical black holes at large redshifts. It is widely known that this idea opens up the possibility of testing the existence of PBHs by observing GW transients at large cosmological distances, even when $f_\pbh \ll 1$. However, it should be noted that observing black hole mergers at high redshifts will be challenging for GW astronomy in the near future.

\bibliography{bib.bib}

\end{document}